\documentclass[aps,prl,twocolumn,showpacs,superscriptaddress,nofootinbib,nobibnotes]{revtex4-1}

\usepackage{breqn}
\usepackage{graphicx}
\usepackage{dcolumn}   
\usepackage{bm}        
\usepackage{amssymb}   
\usepackage{soul}
\usepackage{gensymb}
\usepackage[normalem]{ulem}

\usepackage[breaklinks=true,colorlinks=true,linkcolor=blue,urlcolor=blue,citecolor=blue]{hyperref}

\usepackage{float}

\begin{document}

\title{First-principles based plasma profile predictions for optimized stellarators}

\author{A.~Ba\~n\'on~Navarro}
\affiliation{Max Planck Institute for Plasma Physics Boltzmannstr 2 85748 Garching Germany}
\author{A.~Di~Siena} 
\affiliation{Max Planck Institute for Plasma Physics Boltzmannstr 2 85748 Garching Germany}
\author{J.~L.~Velasco} 
\affiliation{Laboratorio Nacional de Fusi\'{o}n CIEMAT 28040 Madrid Spain}
\author{F.~Wilms} 
\affiliation{Max Planck Institute for Plasma Physics Boltzmannstr 2 85748 Garching Germany}
\author{G.~Merlo}
\affiliation{The University of Texas at Austin 201 E 24th St 78712 Austin Texas USA}
\author{T.~Windisch} 
\affiliation{Max Planck Institute for Plasma Physics Wendelsteinstr 1 17491 Greifswald Germany}
\author{L.~L.~LoDestro}
\affiliation{Lawrence Livermore National Laboratory Livermore CA~94550 USA}
\author{J.~B.~Parker}
\affiliation{Lawrence Livermore National Laboratory Livermore CA~94550 USA}
\author{F.~Jenko} 
\affiliation{Max Planck Institute for Plasma Physics Boltzmannstr 2 85748 Garching Germany}

\begin{abstract}

In the present Letter, first-of-its-kind computer simulations predicting plasma profiles for modern optimized stellarators -- while self-consistently retaining neoclassical transport, turbulent transport with 3D effects, and external physical sources -- are presented. These simulations exploit a newly developed coupling framework involving the global gyrokinetic turbulence code GENE-3D, the neoclassical transport code KNOSOS, and the 1D transport solver TANGO. This framework is used to analyze the recently observed degradation of energy confinement in electron-heated plasmas in the Wendelstein 7-X stellarator, where the central ion temperature was "clamped" to $T_i \approx 1.5$ keV regardless of the external heating power. By performing first-principles based simulations, we provide key evidence to understand this effect, namely the inefficient thermal coupling between electrons and ions in a turbulence-dominated regime, which is exacerbated by the large $T_e/T_i$ ratios, and show that a more efficient ion heat source, such as direct ion heating, will increase the on-axis ion temperature. This work paves the way towards the use of high-fidelity models for the development of the next generation of stellarators, in which neoclassical and turbulent transport are optimized simultaneously.

\end{abstract}


\maketitle


{\em Introduction.} With the systematic reduction of neoclassical transport in modern optimized stellarators like Wendelstein 7-X (W7-X), the magnetic confinement is now turbulence-limited~[\onlinecite{Pedersen_2018, wolf_2019, Klinger_2019}], like in tokamaks. A prominent example of the consequences of this fact was found during the first experimental campaign of W7-X. Here, the observed energy confinement in electron-heated plasmas stayed way below the expectations~[\onlinecite{turkin_2011}]. In fact, the on-axis ion temperature appeared clamped to $T_i \sim 1.5$ keV, regardless of adjustments in the heating power, plasma density, and magnetic configuration. A possible explanation was proposed in Ref.~[\onlinecite{Beurskens_2021}], attributing
this effect to the employed auxiliary heating scheme, i.e., electron cyclotron resonance heating (ECRH).  Specifically, ECRH induces large electron-to-ion temperature ratios, thus possibly enhancing the ion temperature gradient (ITG) driven turbulent transport and resulting in the clamping of the ion temperature profile~[\onlinecite{Beurskens_2021}].

Although some numerical evidence supporting this physical interpretation -- based on gyrokinetic simulations in flux-tube geometry -- was presented in Ref.~[\onlinecite{Beurskens_2021}], the potentially important roles played by 3D effects, neoclassical transport, the radial electric field, and the implications of the limited ion heating from the electron-to-ion exchange power were not explored.
In addition, it remained unclear how the electron-to-ion temperature ratios might affect plasma profile evolution over the entire plasma volume. This is due to the lack of validated transport models in the stellarator community able to correctly capture the combined effect of turbulent transport and external sources. While reduced transport models fail to describe turbulence in complex 3D geometries~[\onlinecite{turkin_2011}], high-fidelity gyrokinetic codes are still limited to short time scales~[\onlinecite{Maurer2020, sanchez_2020, cole_2020, wang_2020}], thus not allowing to consider the evolution of the equilibrium profiles. These hard limits of the presently available modeling tools did not allow reliable prediction of plasma profile evolution and performances in modern-optimized stellarators resulting in the unexpected confinement degradation (ion temperature clamping) as shown in W7-X experiments.

In this Letter, we fill this gap by presenting a new integrated modelling tool, fully developed for optimized stellarator devices, where turbulence, neoclassical transport, the radial electric field, and 3D effects are all retained self-consistently. By applying this tool, we simulate -- for the first time -- the ion temperature profile evolution for electron-heated plasmas in W7-X and reproduce the ion temperature clamping observed in the experiments. In addition, we demonstrate that the interpretation of the clamping presented in Ref.~[\onlinecite{Beurskens_2021}] can only partially address the experimental observations. More precisely, we show that the $T_i$ clamping is determined by inefficient thermal coupling between electrons and ions of electron-heated plasmas in a turbulence-dominated regime, which is exacerbated by the increased turbulence stiffness as the $T_e/T_i$ increases, with the consequent enhancement of ion-scale turbulence. The combination of these effects causes a flattening of the ion temperature profile, severely limiting the performance of these discharges. Finally, we show that a more efficient ion heating source, e.g., direct ion heating, will increase the on-axis temperature.  This work represents first steps towards the development of advanced scenarios in modern optimized stellarators using a high-fidelity turbulence model, and it paves the way for the design of the next generation of stellarator devices, where neoclassical and turbulent transport are optimized simultaneously.

\begin{figure}[!tb]
\begin{center}
\begin{tabular} {c}
\includegraphics[width=0.48\textwidth]{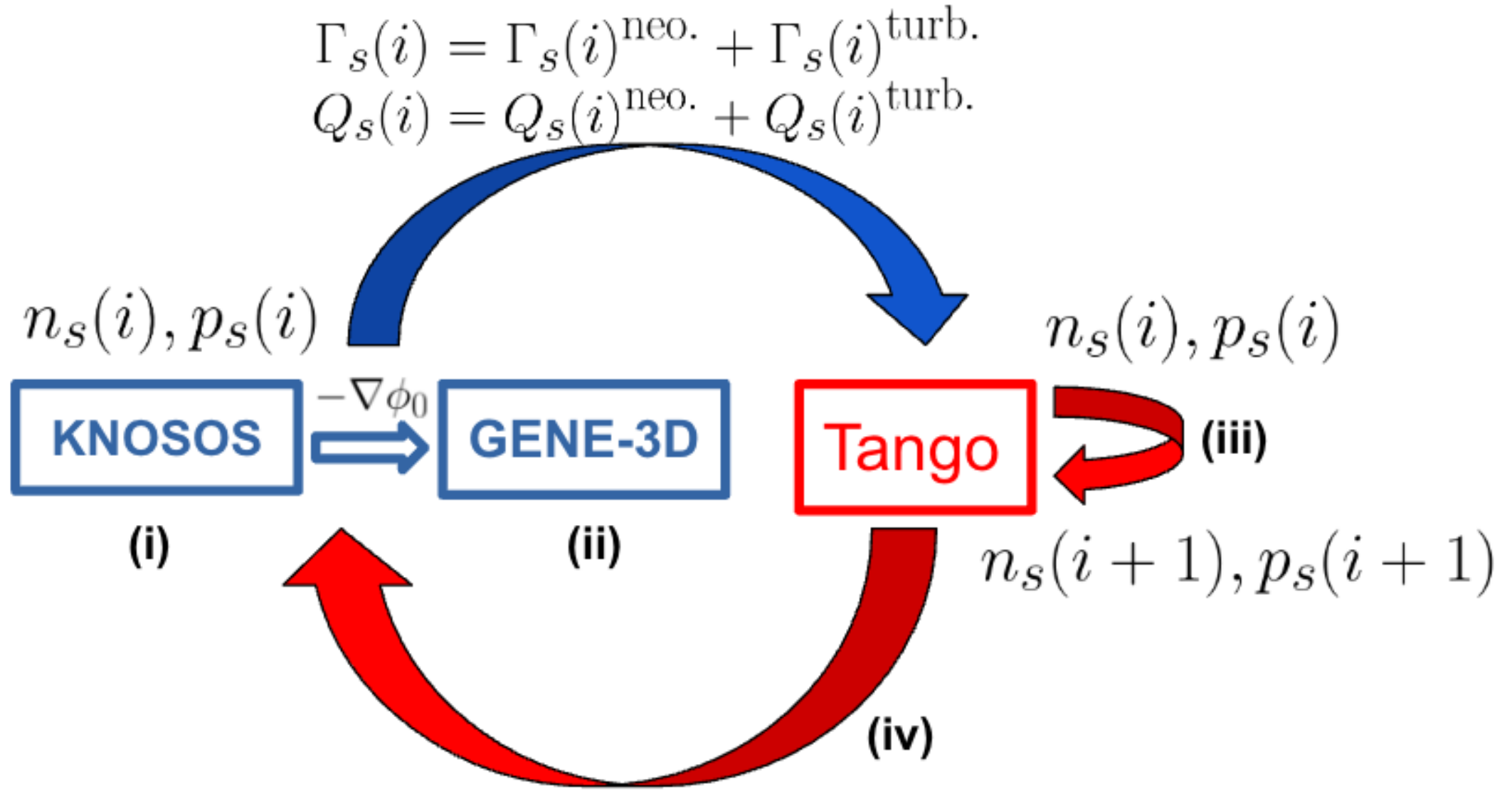} 
\par
\end{tabular}
\end{center}
\vspace{-0.7cm}
\caption{GENE-3D-KNOSOS-Tango workflow diagram: (i) With an initial set of density and pressure profiles for each species ($n_s(i),  p_s(i)$), KNOSOS evaluates neoclassical particle and heat fluxes $(\Gamma_s(i)^{\rm neo.}, Q_s(i)^{\rm neo.})$, and the background radial electric field  $(-\nabla \phi_0)$. (ii) GENE-3D computes the turbulent particle and heat transport $(\Gamma_s(i)^{\rm turb.}, Q_s(i)^{\rm turb.})$ over a number of microscopic time steps using these profiles and the background radial electric field computed by KNOSOS. (iii) Tango evaluates new plasma profiles that are consistent with both the overall (neoclassical and turbulent) fluxes as well as the particle and heat sources. (iv) The process is repeated until a flux equilibrium is reached.}
\label{fig:fig1a}
\end{figure}

{\em Model description.} The integrated, first-principles based transport model involves the global gyrokinetic turbulence code GENE-3D~[\onlinecite{Maurer2020},\onlinecite{wilms_2021}], the neoclassical transport code KNOSOS~[\onlinecite{Velasco_2020},\onlinecite{velasco_2021}], and the 1D transport code TANGO~[\onlinecite{Parker_2018_a},\onlinecite{Parker_2018}]. To achieve self-consistency, the three codes are applied in a continuous loop, see Fig.~\ref{fig:fig1a} for a schematic representation of the workflow. First, KNOSOS evaluates the neoclassical fluxes and the background radial electric field for a given set of density/temperature profiles and a given magnetic equilibrium. On the basis of these plasma profiles and the background radial electric field as computed by KNOSOS, GENE-3D evaluates the turbulent transport over a set of microscopic time steps. Finally, TANGO evaluates new plasma profiles  consistent with both the total (neoclassical and turbulent) fluxes and the particle and heat sources. This process is repeated iteratively until a flux equilibrium is reached.

This high-fidelity model has the unique capability of capturing a wide range of possibly relevant effects, including internal transport barriers~[\onlinecite{garbet_2002, Strugarek_2013, disiena_2021}], turbulence spreading~[\onlinecite{Garbet_1994, Diamond_PPCF2005, lin_2004}], transport avalanches~[\onlinecite{pradalier_2010, mcmillan_2010, sarazin_2000}], field-line dependencies~[\onlinecite{Xanthopoulos_2014}, \onlinecite{disiena_2020}], neoclassical effects~[\onlinecite{Watanabe_2011, Riemann_2016, pavlos_2020}], and external particle/heat sources. The inter-code coupling exploits the time scale separation between turbulence and transport phenomena, and benefits from running each code only on their natural time scales ~[\onlinecite{canday_2019}, \onlinecite{barnes_2010}]. This approach is able to drastically reduce the computational cost for evolving plasma profiles up to the energy confinement time~[\onlinecite{gene_tango_paper}].

\begin{figure}[!tb]
\begin{center}
\begin{tabular} {c}
\includegraphics[width=0.48\textwidth]{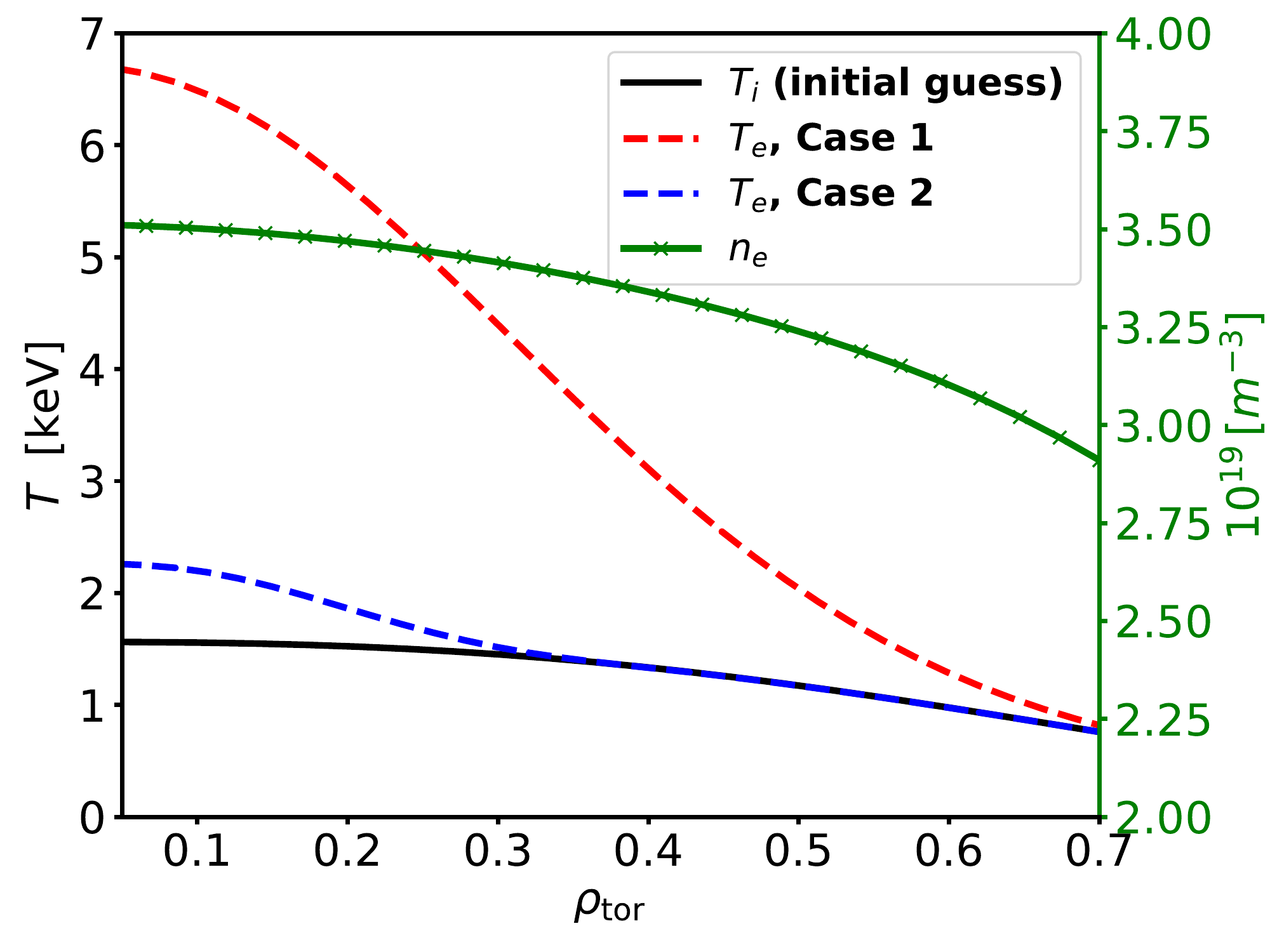} 
\par
\end{tabular}
\end{center}
\vspace{-0.7cm}
\caption{Temperature and density profiles.
The central $T_{e,0}$ is varied from roughly $7$ keV (Case 1) to  $2$ keV (Case 2).
The same ion temperature profile with $T_{i,0} = 1.5$  keV (initial guess) and density profile with $n_{e,0} = 3.5 \,\cdot 10^{19} \, m^{-3}$ is chosen for both cases.
}
\label{fig:fig1}
\end{figure}

{\em Simulation setup.} The numerical simulations presented in this Letter are performed for hydrogen plasmas heated by ECRH in the standard W7-X configuration [\onlinecite{Dinklage_2018}]. The GENE-3D simulations are collisionless and assume an adiabatic electron response. As a result, electron heat and particle fluxes are neglected, and the $T_e$ and $n_e$ profiles are fixed in time. This is done to significantly lower the computational cost of the simulations. These profiles are inspired by the typical plasma profiles observed in ECRH-dominated W7-X plasmas~[\onlinecite{Beurskens_2021}, \onlinecite{ Carralero_2021}]. While the on-axis plasma density is fixed to $n_{e,0} = 3.5 \,\cdot 10^{19} \, m^{-3}$ with relative flat gradient, two different electron temperature profiles are considered: one with $T_{e,0} = 7$ keV (Case 1) and another with $T_{e,0} =2$ keV (Case 2). The initial guess for the ion temperature profile is taken from the experimental measurements of W7-X with a clamped on-axis temperature of $T_{i,0}=1.5$ keV (see Fig.~\ref{fig:fig1}).
GENE-3D is run using the gradient-driven approach~[\onlinecite{goerler_2011}], which employs a Krook-type heat source to maintain the plasma profiles close to the ones provided by TANGO at each iteration. Zero Dirichlet boundary conditions are employed with buffer zones covering $5-10\%$ of the radial domain where a Krook damping operator is applied to avoid nonphysical profile variations close to the boundaries. The numerical resolution in $(x,y,z,v_{\parallel},\mu)$ space used in GENE-3D is $(192 \times 256 \times 128 \times 32 \times 8)$. Here, $x = \rho_{\rm tor}$ is a radial coordinate based on the toroidal flux $\Phi, \rho_{\rm tor}= \sqrt{\Phi/\Phi_{\rm LCFS}}$, normalized by its value at the last close flux surface; $y$ is a coordinate along the binormal direction; $z$ the coordinate along the magnetic field line; $v_{\parallel}$ is the parallel velocity and $\mu$ is the magnetic moment. The radial simulation domain is $\rho_{\rm tor} = [0.05-0.7]$. In the binormal direction, we exploit the five-fold symmetry of W7-X and consider one-fifth of the toroidal domain. TANGO solves the ion pressure equation using $27$ radial grid points, and the electron-to-ion exchange acts as the ion heating source~[\onlinecite{Huba_2013}]. KNOSOS uses the same radial grid as TANGO to calculate the neoclassical heat fluxes and background radial electric field using the ambipolarity constraint~[\onlinecite{velasco_2021}].

\begin{figure}[!tb]
\begin{center}
\begin{tabular} {c c}
\includegraphics[width=0.48\textwidth]{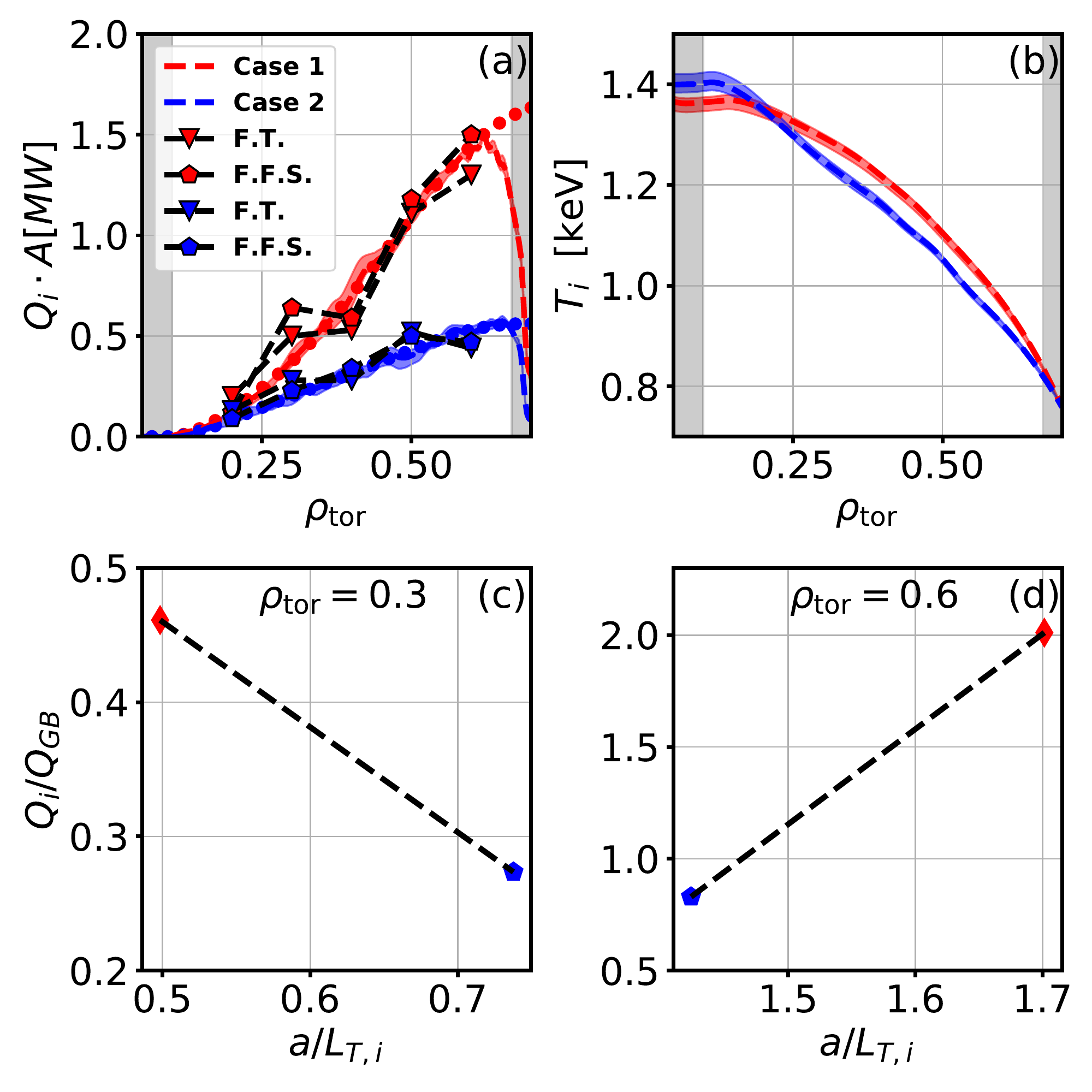} 
\par
\end{tabular}
\end{center}
\vspace{-0.7cm}
\caption{(a) Comparison of the ion heat flux computed by the integrated model for Case 1 (red) and Case 2 (blue). The shaded areas represent the fluctuations of the fluxes over the last five iterations.  The gray areas denote the buffer regions employed in the GENE-3D simulations, and the circles the volume integral of the heat sources (from the electron-to-ion exchange power).  
The markers denote the results of local full-flux-surface (labelled as F.F.S. and with pentagons) and bean-shaped flux-tube (F.T. with triangles) simulations obtained with GENE-3D using the converged profiles from TANGO. (b) Resulting ion temperature profiles for both cases. Dependency of the normalized total ion heat flux $Q_i/Q_{GB}$ at (c) $\rho_{\rm tor} =0.3$ and (d) at $\rho_{\rm tor} = 0.6$ on the normalized temperature gradient  $a/L_{T,i}$. Here, $Q_{GB} = v_i n_i T_i (\rho_i/ a)^2$, with the ion thermal velocity $v_i = \sqrt{T_i / m_i}$ and the ion Larmor radius $\rho_i = v_i / \Omega_i$, where $\Omega_i = (q_i B_0) / (m_i c)$ is the ion gyrofrequency, $m_i$ the ion mass, $q_i$ the ion charge, $c$ the speed of light and $B_0$ the equilibrium magnetic field on-axis.
}
\label{fig:fig2}
\end{figure}

{\em Ion heating in ECRH plasmas.} We now apply GENE-3D-KNOSOS-TANGO to study the ion temperature clamping effect observed at W7-X. To cover different electron-to-ion heating powers and electron-to-ion temperature ratios, we perform simulations considering the different electron temperature profiles introduced above. For each of these cases, GENE-3D-KNOSOS-TANGO is run until it reaches a steady-state solution. The converged results are shown in Fig.~\ref{fig:fig2}a where the ion (turbulent and neoclassical) heat fluxes computed by GENE-3D-KNOSOS match the volume integral of the electron-to-ion exchange power for the two cases considered.

Interestingly, qualitatively, both simulations lead to the same ion temperature profile (Fig.~\ref{fig:fig2}b), despite the large differences (almost a factor of 3 at $\rho_{\rm tor} = 0.7$) in ion heating. In particular, the central ion temperature is clamped to about $1.35$ keV. These findings are consistent with the experimental observations~[\onlinecite{Beurskens_2021}], showing a negligible impact of the ion heating power from the electron-to-ion exchange power on the on-axis ion temperature.
When looking at the normalized gradient ($a/L_{T,i} = - a \,d \ln (T_i) / {d \rho_{\rm tor}}$ with $a$ being the minor radius) dependence on the ion heating power, we notice a complex behavior between the inner and outer core positions. While the normalized gradient  increases with ion heating (in gyro-Bohm units) at $\rho_{\rm tor} = 0.6$, it diminishes by a similar amount at $\rho_{\rm tor} = 0.3$. This is shown in Fig.~\ref{fig:fig2}c-d for the two Cases studied here.  This dependency of the normalized gradient  on the ion heating power is also observed experimentally [\onlinecite{Beurskens_2021}].

\begin{figure}[!tb]
\begin{center}
\begin{tabular} {c}
\includegraphics[width=0.48\textwidth]{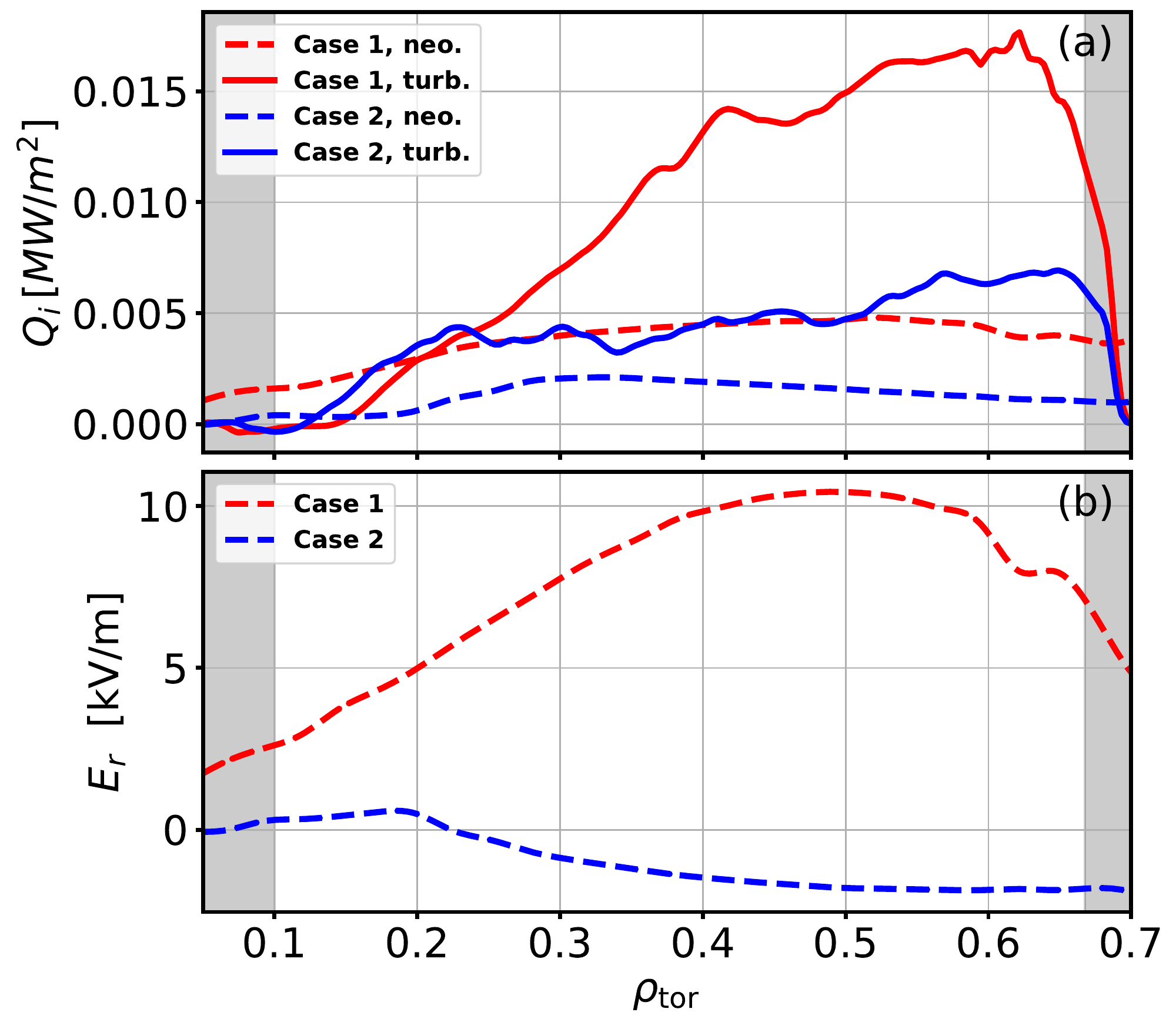} 
\par
\end{tabular}
\end{center}
\vspace{-0.7cm}
\caption{ (a) Contribution of the neoclassical (neo.) and turbulent (turb.) transport to the total heat flux and (b) background radial electric field calculated by KNOSOS for both cases.}
\label{fig:fig2a}
\end{figure}

Furthermore, in Fig.~\ref{fig:fig2a}a, we illustrate the contribution of neoclassical transport and turbulent transport to the total heat flux for both cases. In the majority of the radial domain, neoclassical transport is subdominant compared to turbulent heat transport. Neoclassical transport becomes similar to turbulent transport only in the inner core (up to approximately 0.2 $\rho_{\rm tor}$), where ion profiles flatten and turbulent transport reduces dramatically. Finally, we compare the radial electric field calculated by KNOSOS for both cases in Fig.~\ref{fig:fig2a}b. There are significant differences between the two scenarios. In particular, Case 1 has an electron (positive) root due to the high electron temperature, whereas Case 2 has an ion (positive) root in the inner core (up to $\rho_{\rm tor} \approx 0.2$) and an ion (positive) root in the remaining radial domain. 
%


{\em Understanding the $T_i$ clamping in ECRH plasmas}
Various effects might be considered as the cause of the ion temperature clamping observed in W7-X, including neoclassical effects~[\onlinecite{Watanabe_2011, Riemann_2016, pavlos_2020}], 3D~[\onlinecite{mcmillan_2010, sarazin_2000, Xanthopoulos_2014}] and $T_e/T_i$~[\onlinecite{Beurskens_2021}] effects on turbulence, as well the limited ion heating from the electron-to-ion exchange power.

\begin{figure}[!t]
\begin{center}
\begin{tabular} {c}
\includegraphics[width=0.48\textwidth]{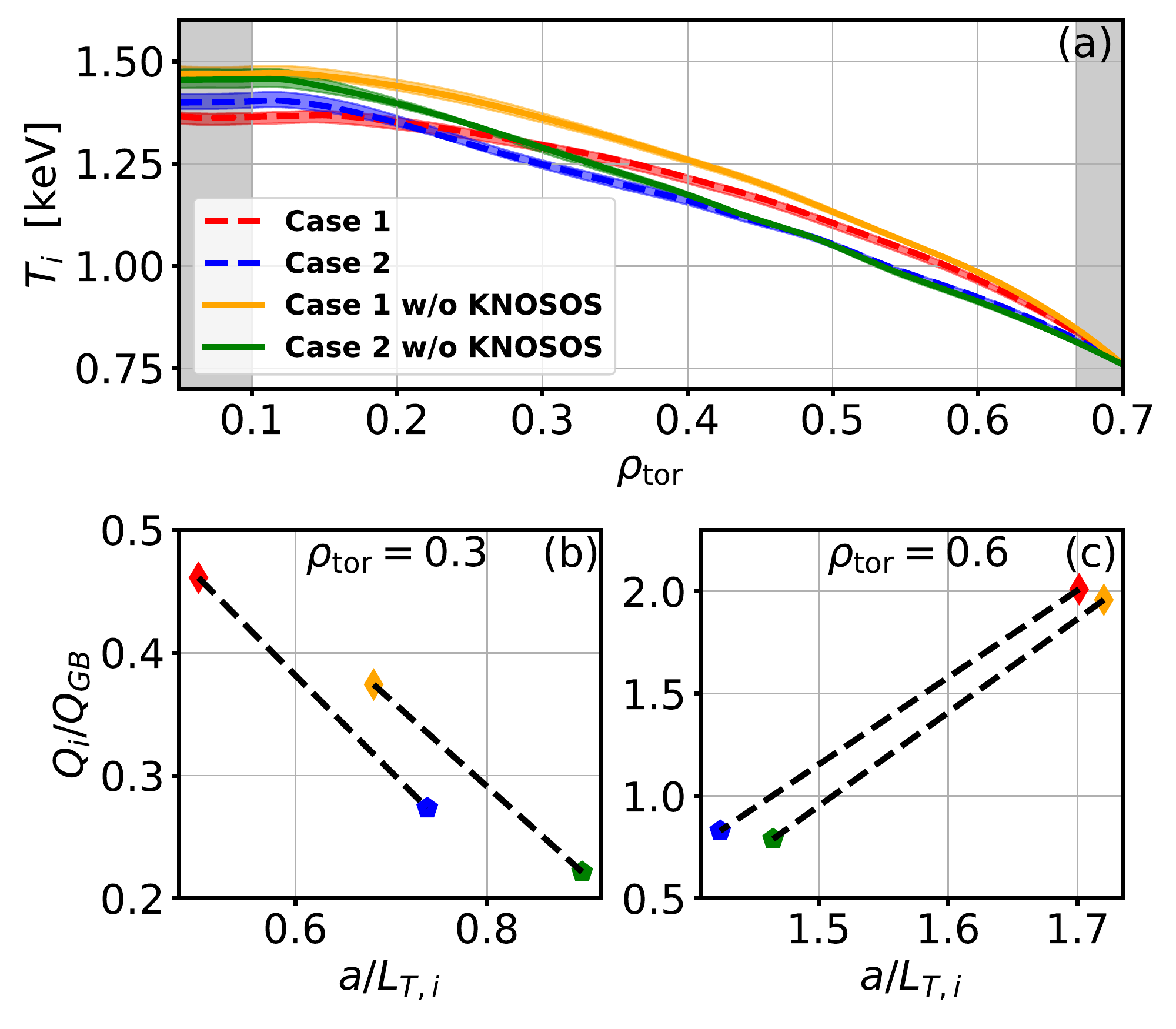}  
\par
\end{tabular}
\end{center}
\vspace{-0.7cm}
\caption{(a) Ion temperature profile for the cases with and without KNOSOS from the GENE-3D-TANGO loop. 
Dependency of the normalized ion heat flux $Q_i/Q_{GB}$ at (b) $\rho_{\rm tor} =0.3$ and (c) at $\rho_{\rm tor} = 0.6$ on the normalized temperature gradient  $a/L_{T,i}$. 
}
\label{fig:fig3}
\end{figure}

We begin our analysis by investigating the role played by the neoclassical transport and the radial electric field in the GENE-3D-KNOSOS-TANGO computations for the two scenarios discussed above. More specifically, we perform simulations removing KNOSOS from the loop. Therefore, the background radial electric field is not retained in the GENE-3D calculations and the TANGO profiles are evaluated neglecting the neoclassical contribution to the overall fluxes. The resulting ion temperature profiles are shown in Fig.~\ref{fig:fig3}a. No significant impact on the central ion temperature is observed, which -- despite undergoing a mild peaking in the cases without KNOSOS -- remains clamped to $T_i \approx 1.35-1.45$ keV in all cases considered. Additionally, the inverse response of the normalized gradients to the ion heat flux at $\rho_{\rm tor} = 0.3$ is still present (Fig.~\ref{fig:fig3}b). Therefore, neoclassical transport and background radial electric field effects cannot explain the ion temperature clamping observed in W7-X and will be neglected for simplicity in the remainder of this Letter.

Other effects that might lead to $T_i$ clamping are 3D effects, namely radially global and magnetic field-line dependencies. In the following, we assess their role in affecting the $T_i$ evolution by performing local full-flux-surface and bean-shaped flux-tube simulations with GENE-3D at different radial locations. This is possible thanks to the unique capabilities of GENE-3D, which can be run in either flux-tube, full flux-surface, or radially global domains.   For this analysis, we used the converged profiles of Fig.~\ref{fig:fig3}a. The results are summarized in Fig.~\ref{fig:fig2}a, showing an excellent agreement between the global and local heat fluxes for both Case 1 and Case 2, thus suggesting that a local GENE-3D-TANGO framework would lead to the same $T_i$ profiles. Therefore, the ion temperature clamping cannot be not induced by 3D effects. It is worth mentioning that 3D effects might play a non-negligible role in other scenarios with smaller $1/\rho^{*}$ (here $1/\rho^{*}  = a/ \rho_i \gtrsim 350$)~[\onlinecite{Xanthopoulos_2014}].

\begin{figure}[!t]
\begin{center}
\begin{tabular} {c}
\includegraphics[width=0.48\textwidth]{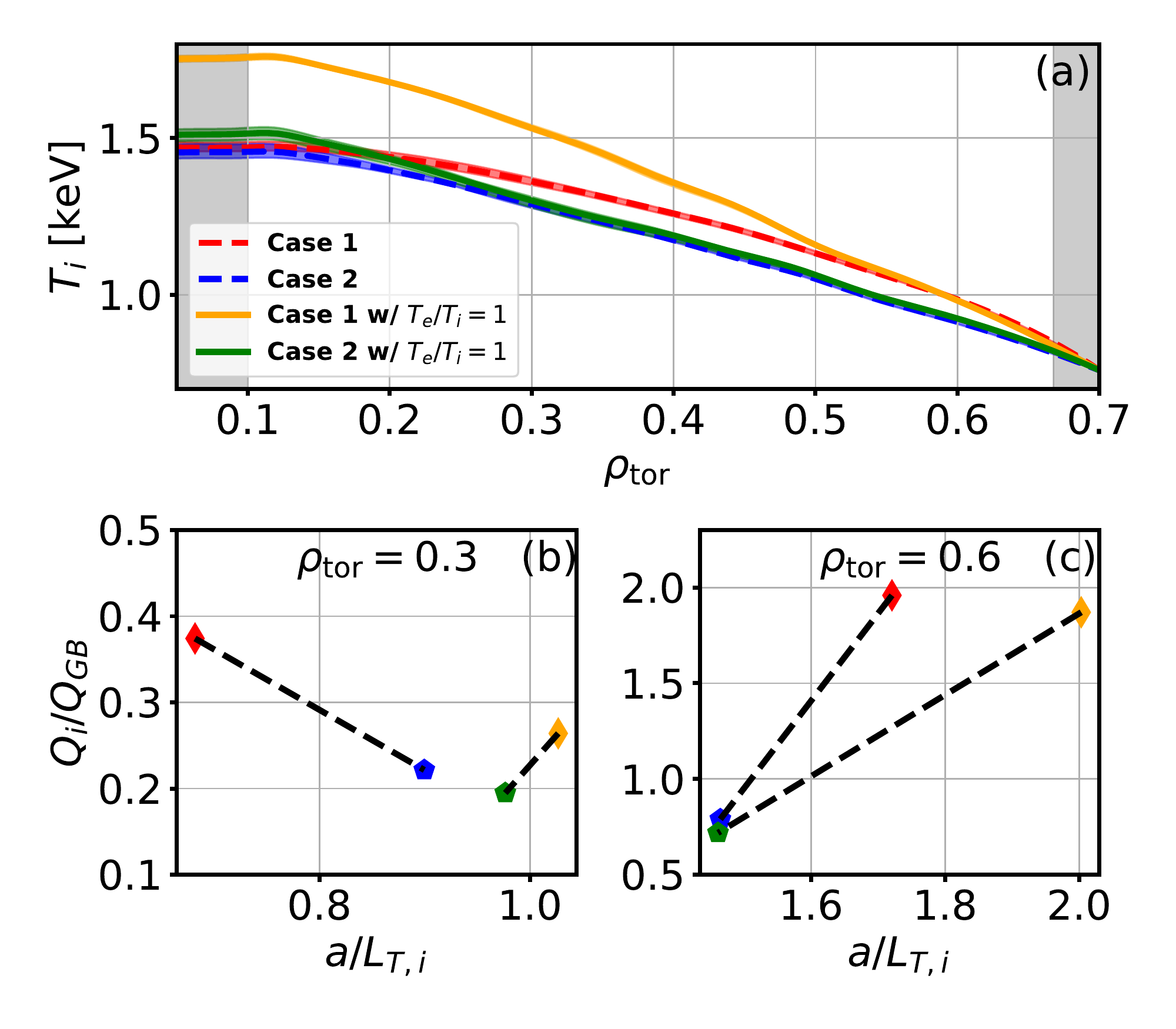} 
\par
\end{tabular}
\end{center}
\vspace{-0.7cm}
\caption{(a) Ion temperature profile comparing Case 1 and 2 with the cases where $T_e/T_i = 1$ is forced in the GENE-3D field equation but keeping its finite value in TANGO. Dependency of the normalized ion heat flux $Q_i/Q_{GB}$ at (b) $\rho_{\rm tor} =0.3$ and (c) at $\rho_{\rm tor} = 0.6$ on the normalized temperature gradient  $a/L_{T,i}$ for those cases.}
\label{fig:fig4}
\end{figure}

We now focus on how plasma turbulence -- determining the plasma profiles -- is influenced by the electron-to-ion temperature ratio. This is done with GENE-3D-TANGO by forcing $T_e = T_i$ in the GENE-3D field equation but keeping its actual value in TANGO. By doing so, the ratio $T_e/T_i$ will not affect ITG turbulence but will be considered in TANGO to evaluate the electron-to-ion power exchange. With this specific setup, we observe a $20 \%$ increase in Fig.~\ref{fig:fig4}a of the on-axis ion temperature for Case 1, while it is largely unaffected for Case 2. This negligible role played by $T_e/T_i$ on the ion profile for Case 2 is due to the specific choice of the electron temperature profile. As shown in Fig.~\ref{fig:fig1}, the electron temperature already follows the ion profile, thus rendering any $T_e/T_i$ effects negligible. Interestingly, the inverse response of normalized gradient  to the ion heating at the inner core is inverted. This is illustrated in Figs.~\ref{fig:fig4}b and c, where $a/L_{T,i}$ is observed to increase with the normalized ion heat flux not only at $\rho_{\rm tor} = 0.6$, but also at $\rho_{\rm tor} = 0.3$ when $T_e/T_i=1$ in GENE-3D.

Finally, we assess the impact of the limited ion heating from the electron-to-ion exchange power on the clamping. This is done by performing GENE-3D-TANGO simulations, including an external ion heating source in the modelling, and progressively increasing its amplitude from $1$ MW to $10$ MW. Its shape is described by a Gaussian function centered at $\rho_{\rm tor} = 0.2$. This is a good approximation for ion-cyclotron-resonance heating (ICRH) sources~[\onlinecite{Brambilla_2006}].  Given the significant role played by $T_e/T_i$, the dependency of the central ion temperature on the injected ion power is studied by fixing $T_e/T_i$ to the values characterizing the two cases presented previously. This is achieved by rescaling $T_e$ accordingly at each TANGO iteration to keep $T_e/T_i$ constant. 

\begin{figure}[!t]
\begin{center}
\begin{tabular} {c}
\includegraphics[width=0.48\textwidth]{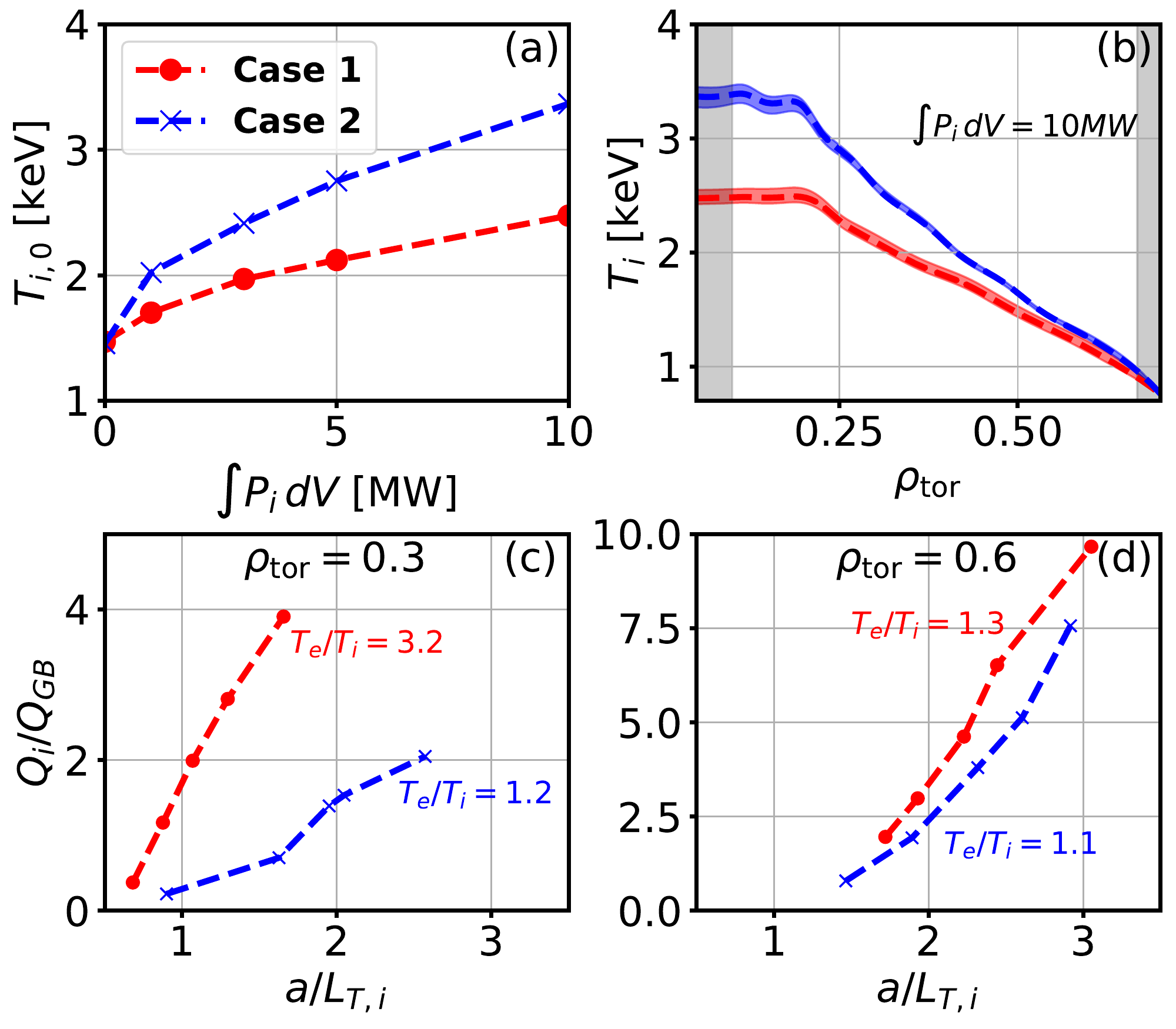} 
\par
\end{tabular}
\end{center}
\vspace{-0.7cm}
\caption{(a) Response of $T_i$ on-axis with the injected ion heating power, where the $0$ value corresponds to the case where there is not a direct ion heating source and the ions are only heated via the electron-to-ion exchange power. (b) Resulting ion temperature profiles for $\int P_i \rm{d}V= 10$ MW of injected  ion power. Dependency of the normalized ion heat flux $Q_i/Q_{GB}$ at (c) $\rho_{\rm tor} =0.3$ and (d) at $\rho_{\rm tor} = 0.6$ on the normalized temperature gradient  $a/L_{T,i}$ for those cases.
}
\label{fig:fig5}
\end{figure}

The results are summarized in Fig.~\ref{fig:fig5}a. Interestingly, we observe that heating the ions directly increases the central ion temperature for both cases. This is increasingly effective --at a fixed heating power -- when $T_e/T_i \approx 1$ is ensured. In particular, Fig.~\ref{fig:fig5}a shows that the central ion temperature raises roughly by a factor of $2.3$ for Case 2, leading to a central value of approximately $3.5$ keV for an injected power of $10$ MW. For the same power, Case 1 -- having a large $T_e/T_i$ in the inner radius -- shows a $1.6$ factor increase, reaching only $T_{i,0} \approx 2.5$ keV (see Fig.~\ref{fig:fig5}b for a closer comparison). Furthermore, we observe a different behavior with the radial location when looking at the heat flux dependence with the normalized temperature gradient  (transport stiffness). At $\rho_{\rm tor} =0.6$, $T_e/T_i \approx 1$ for both cases and Fig.~\ref{fig:fig5}d shows a similar stiffness (defined as the slope of Fig.~\ref{fig:fig5}d). This changes in the inner core locations, where $T_e > T_i$ for Case 1, leading to an increased stiffness at $\rho_{\rm tor} =0.3$ with respect to Case 2 which keeps $T_e/T_i \approx 1$, as shown in Fig.~\ref{fig:fig5}c. Therefore, for a fixed injected power, the largest logarithmic temperature gradient 
is reached when the ion temperature closely follows its electron counterpart. 


{\em Discussion and conclusions.} The first-of-their-kind integrated simulations presented in this Letter are able to reproduce and explain the clamping of the central ion temperature observed in W7-X. We demonstrated that $T_i$ clamping is induced by inefficient thermal coupling between electrons and ions in electron-heated plasmas in a turbulence-dominated regime, which is exacerbated by an increase in turbulence stiffness when $T_e/T_i$ rises, resulting in an increase in ion-scale turbulence. The combination of these effects causes the ion temperature profile to flatten, greatly restricting the performance of these discharges. In fact, a more efficient ion heat source, such as direct ion heating, raises the on-axis temperature. When $T_e/T_i \approx 1$ is ensured, this becomes more effective for a given heating power.
Our simulations, on the other hand, show that 3D effects are negligible, and that neoclassical transport and background radial electric influence the on-axis ion temperature by about 10\%.

These results were obtained with the first high-fidelity model in the stellarator community able to compute
plasma profiles due to the combined effects of neoclassical transport, turbulent transport, and external particle/heat sources in 3D magnetic equilibria. This was achieved by coupling a global gyrokinetic code (GENE-3D), a neoclassical code (KNOSOS), and a transport code (TANGO), thus correctly capturing both neoclassical and 3D effects on plasma turbulence.

In future studies, we would aim to include a more thorough plasma description within the simulations. Specifically, we would like to incorporate kinetic electrons, collisions, and electromagnetic effects into GENE-3D such that the temperature and density profiles of electrons can also evolve. The objective is to perform a first-principles validation study for W7-X, which will give us confidence in our model and help with profile prediction and future scenario development in W7-X, as well as explore new optimized stellarator configurations~[\onlinecite{Bader_2021,landreman_2022,clark_2022,edi2023}] and identify the most efficient stellarator design capable of simultaneously optimizing turbulent and neoclassical transport.


{\em Acknowledgements.} The authors would like to express their gratitude to C. Angioni and C. Bourdelle for their insightful discussions and encouragement to consider the role of inefficient thermal coupling between electrons and ions in the  clamping.  We also thank R.~Bilato and G.~Plunk for stimulating discussions, useful suggestions, and comments. This work has been carried out within the frame-work of the EUROfusion Consortium, funded by the European Union via the Euratom Research and Training Programme (Grant Agreement No 101052200 — EURO-fusion). Views and opinions expressed are however those of the author(s) only and do not necessarily reflect those of the European Union or the European Commission.Neither the European Union nor the European Commission can be held responsible for them. Furthermore, numerical simulations were performed at the Marconi Fusion supercomputers at CINECA, Italy.


%

\end{document}